\documentclass{sf2a-conf2013}
\usepackage{graphicx}
\usepackage{hyperref}
\usepackage[]{natbib}  
\usepackage{epstopdf}

\def\BibTeX{{\rm B\kern-.05em{\sc i\kern-.025em b}\kern-.08em
    T\kern-.1667em\lower.7ex\hbox{E}\kern-.125emX}}
\bibpunct{(}{)}{;}{a}{}{,}  

\begin{document}

\TitreGlobal{SF2A 2013}

\title{Dust in the wind I:\\ Spectropolarimetric signatures from disk-born outflows}

\runningtitle{Dust in the wind I}

\author{F. Marin$^*$}\address{Observatoire Astronomique de Strasbourg, Universit\'e de Strasbourg, 
			      CNRS, UMR 7550, 11 rue de l'Universit\'e, 67000 Strasbourg, France\\}
		 \thanks{$^*$ frederic.marin@astro.unistra.fr}

\author{R. W. Goosmann$^1$}

\setcounter{page}{237}

\index{Marin, F.}
\index{Goosmann, R. W.}


\maketitle

\begin{abstract}
In this first research note of a series of two, we conduct optical/UV investigations of the spectropolarimetric signatures 
emerging from the structure of quasars \citep{Elvis2000} applied to a purely theoretical, dusty model. We aim to explore the 
similarities/differences between an absorbing, disk-born outflow and the usual dusty torus that is supposed to hide the 
internal regions of active galactic nuclei (AGN). Using radiative transfer Monte Carlo simulations, we compute the continuum 
polarization signatures emerging from the model setup of \citet{Elvis2000}. We find that a dust-filled outflow produces very 
low amount of wavelength-depend polarization degrees, associated with a photon polarization angle perpendicular to the projected 
symmetry axis of the model. The polarization percentages are ten times lower than what can be produced by a toroidal model, 
with a maximal polarization degree found for intermediate viewing angle (i.e. when the observer's line-of-sight crosses 
the outflowing material). The structure for quasars unsuccessfully blocks the radiation from the central irradiating source and 
shows a spectropolarimetric behavior that cannot be conciliated with observations. Either a new set of morphological parameters 
or different optical thickness must be considered.
\end{abstract}



\section{Introduction}
According to the unified model of radio-quiet AGN \citep{Antonucci1993}, most of the observational differences between Seyfert-like galaxies can be attributed to
an orientation effect. The broad line features, spectroscopically detected in type-1 (pole-on) AGN, disappear in type-2 (edge-on) spectra as the emission region
(the Broad Line Region, BLR) is hidden behind an optically thick, circumnuclear medium \citep{Antonucci1985,Pier1992}. This hypothetical region, responsible 
for the anisotropic radiation pattern of quasi-stellar objects (QSO), was recently confirmed by \citet{Jaffe2004} and \citet{Wittkowski2004}. Using mid-infrared 
interferometric measurements, they spatially resolved the direct emission coming from a torus-like structure in the hearth of NGC~1068. A simple toroidal 
configuration was first postulated to achieve obscuration of the central AGN region \citep{Krolik1986,Krolik1988}, but such a model suffers from 
several hydrodynamic issues that cannot sustain parsec-scale cohesion. More complex models using clumps or filamentary structures are actually considered to 
replace this basic torus model \citep{Nenkova2008a,Nenkova2008b,Nenkova2010,Schartmann2008,Hoenig2010,Heymann2012}.

In its structure for quasars, \citet{Elvis2000} tried, among many other goals, to overcome the need for an obscuring torus around the equatorial plane of 
the central engine. To do so, he supposed that the obscuring medium is not static but results in an outflow, originating close to the supermassive black hole
(SMBH) and its accretion disk that powers the AGN. Such a wind is expected to arise from a narrow range of radii on the accretion disk and is driven by
internal radiation pressure. In \citet{Elvis2000}, the outflow is essentially composed of ionized gas, but the presence of dust originating 
close to the accretion disk remains theoretically possible. According to \citet{Czerny2012}, dust can be formed in accretion disk atmospheres, rises similarly 
to the electronic disk-born wind, but is soon evaporated by the central irradiation source \citep{Elvis2012}. 

In this research note, the first of a series of two, we aim to investigate such a purely theoretical hypothesis. We replaced the ionized matter that is supposed to 
arise from the accretion disk by a dust mixture, and ran {\sc stokes} simulations \citep{Goosmann2007,Marin2012}. We produce polarization spectra that we 
compare to the usual torus configuration used for circumnuclear obscuration.

\section{Model setup}

\begin{figure*}[!t]
\centering
   \includegraphics[trim = 0mm 145mm 5mm 40mm, clip, width=16cm]{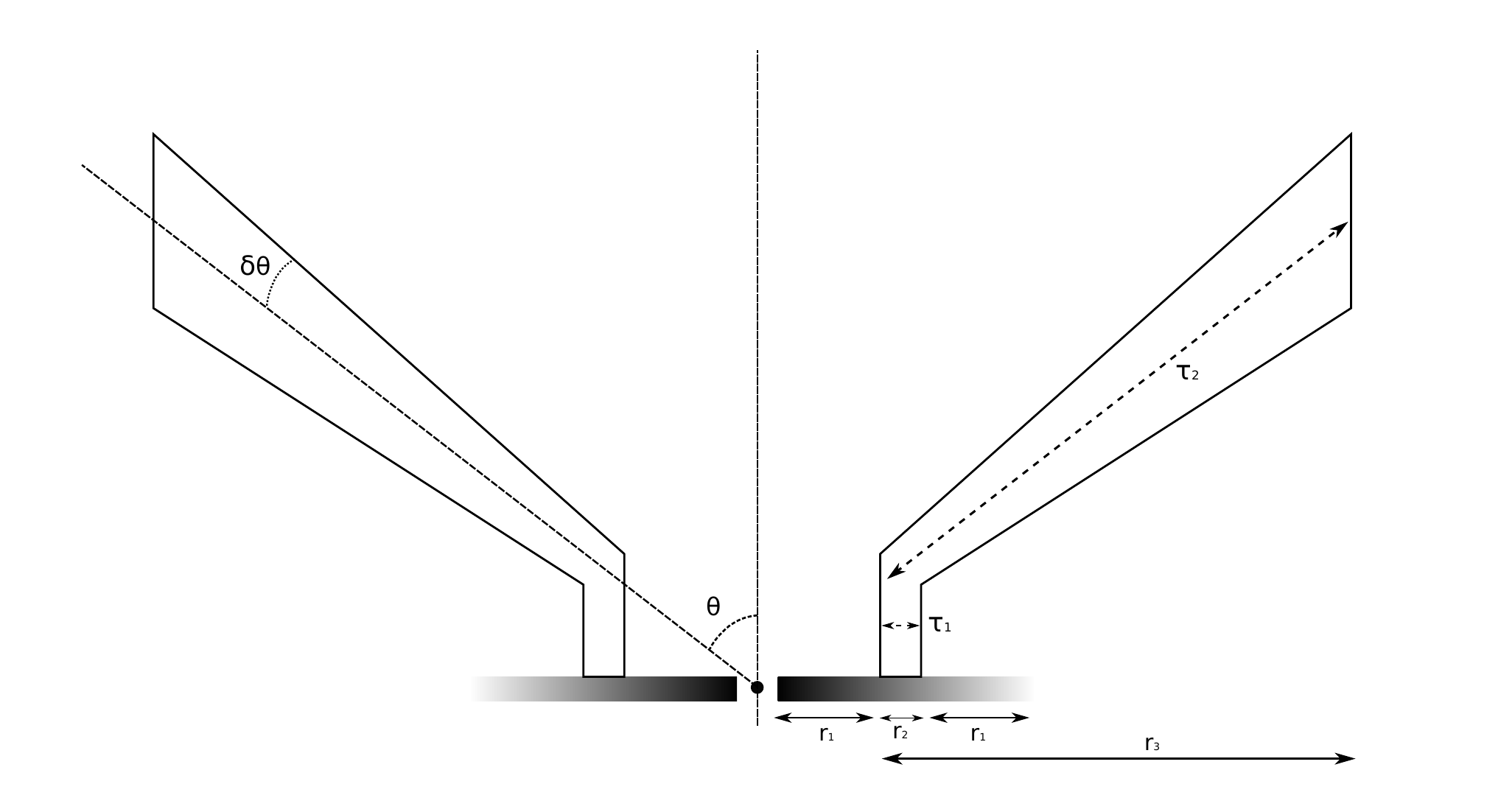}
   \caption{Schematic view of the structure proposed by \citet{Elvis2000} and
            implemented in {\sc stokes}. The outflow arises vertically from the accreting 
	    disk and is bent outward by radiation pressure along an inclination of 60$^\circ$ 
	    relative to the symmetry axis of the model. The half-opening angle of the wind 
	    is equal to 3$^\circ$. The radial optical depths of the wind base and of the 
	    outflowing material are set to $\tau_1$ and $\tau_2$ respectively.}
  \label{Fig1}
\end{figure*}

To explore the polarization signatures of dust in the structure for quasars proposed by \citet{Elvis2000}, we use the radiative transfer Monte Carlo code 
{\sc stokes} to compute the polarization emerging from a continuous medium, whose geometry is summarized in Fig.~\ref{Fig1}. The continuum 
source is defined as an isotropic, disk-like region emitting a unpolarized spectrum with a power-law spectral energy distribution $F_{\rm *}~\propto~\nu^{-\alpha}$ 
and $\alpha = 1$. From the disk, the wind arises vertically at a distance r$_1$ from the model origin. The outflowing column has a width set to r$_2$.
Due to radiation pressure, the wind is bent by an angle $\theta$ relative to the symmetry axis of the model, with a half-opening angle $\delta\theta$, 
and extends up to r$_3$. According to \citet{Elvis2000}, we parametrize our model using r$_1$ = 0.0032~pc (10$^{16}$ cm), r$_2$ = 0.00032~pc, r$_3$ = 0.032 pc, 
$\theta$ = 60$^\circ$ and $\delta\theta$ = 3$^\circ$.

The outflow is filled with a typical ``Milky Way'' dust mixture \citep{Wolf1999}, already used in previous AGN simulations 
\citep{Goosmann2007,Marin2012,Marin2012b,marin2012c}. We opted for an optically thick model, with $\tau_1 \sim$~36 and $\tau_2 \sim$~3600, in order to approximate 
the obscurring impact of a regular dusty torus. This purely theoretical model is not necessarily physical as such a dense absorbing medium may not form or survive so close 
to the accretion disk.

\section{Spectropolarimetric results}

   \begin{figure}[!t]
   \centering
      \includegraphics[trim = 10mm 0mm 0mm 0mm, clip, width=10cm]{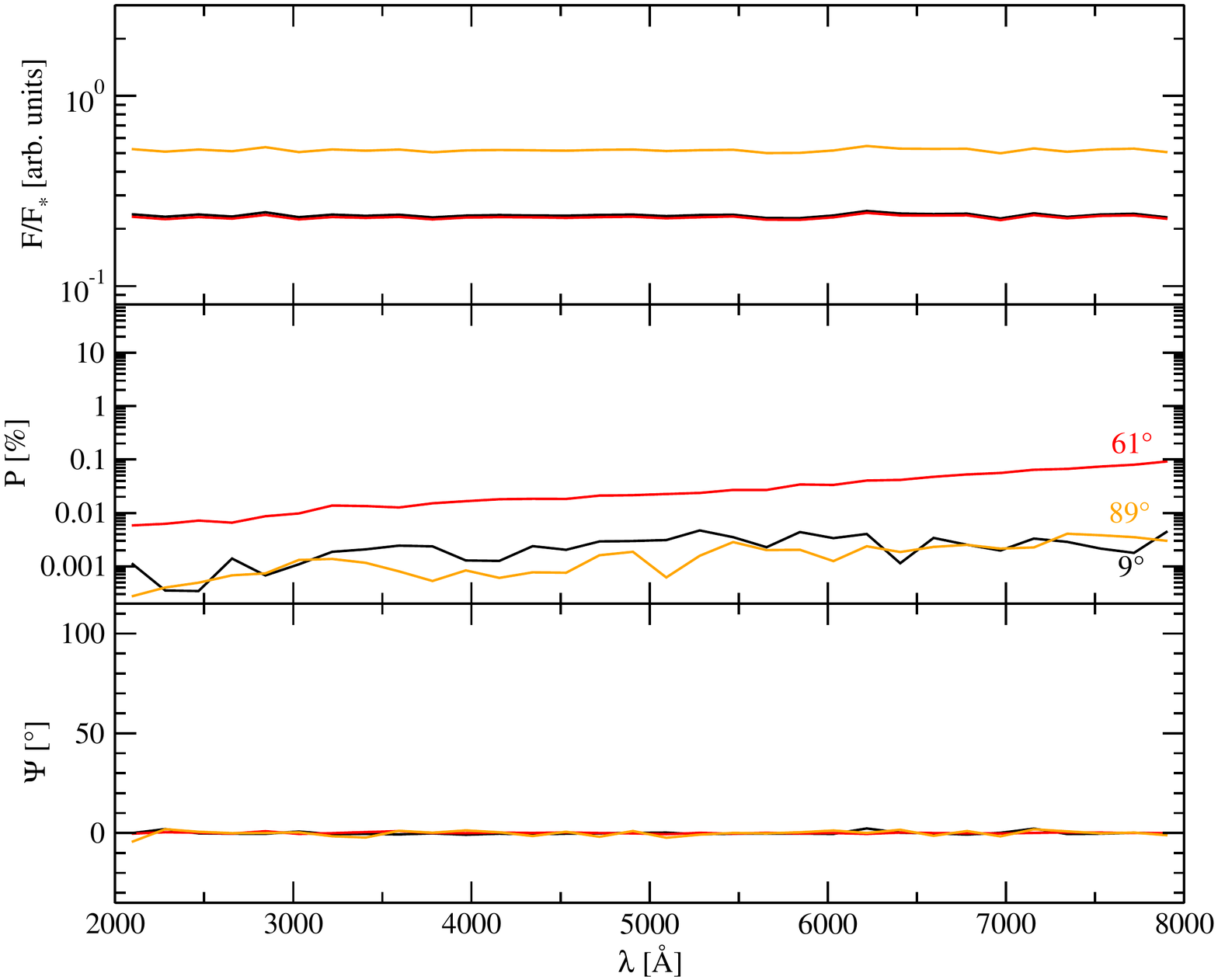}
      \caption{Modeling a dust-filled structure for quasars outflow
	       using a uniform model such as in Fig.~\ref{Fig1},
	       seen at different viewing inclinations, \textit{i} 
	       (black : 9$^\circ$ -- red : 61$^\circ$ -- orange : 89$^\circ$).
	       \textit{Top}: the fraction, $F$/$F_{\rm *}$ of the central flux;
	       \textit{middle}: polarization, \textit{P};
	       \textit{bottom}: polarization angle, $\psi$.}
     \label{Fig2}
   \end{figure}

The spectropolarimetric signatures of the dusty outflows are presented in Fig.~\ref{Fig2}. The fraction $F$/$F_{\rm *}$ of the central flux is 
wavelength-independent for all viewing angles, but the normalization of $F$/$F_{\rm *}$ in polar and equatorial inclinations is different. At $i$ = 9$^\circ$,
most of the flux comes from the unobscured central source. However, photons reprocessing on the dust funnel are mostly absorbed due to 1)
the Mie scattering phase function, promoting forward and backward scattering, and 2) the albedo of the absorbing dust. Across the outflowing gas
($i$ = 61$^\circ$) absorption is most efficient and $F$/$F_{\rm *}$ small. Along the equatorial plane ($i$ = 89$^\circ$), $F$/$F_{\rm *}$ is found to be 
higher than expected from toroidal circumnuclear, dusty regions \citep{Goosmann2007,Marin2011,Marin2012}. The low optical depth of the wind column, 
associated with forward scattering dominated reprocessing, tend to increase the escape probability, resulting in a larger flux detected at edge-on 
inclinations.

The combination of an extended, disk-like radiating source and a small optical depth in the poloidal direction of the outflow increases the escape probability 
and is responsible for relatively high polarization degree at lines-of-sight passing through the dust medium ($P <$ 0.1~\% for $i$~=~61$^\circ$). Due to 
dilution by the extended emitting source and absorption by the dusty wind, the degree of polarization at polar ($i \sim$ 9$^\circ$) and equatorial ($ i\sim$~89$^\circ$)
angles is lower ($P <$ 0.01~\%). We also observe that the percentage of polarization is no longer wavelength-independent and rises towards the red part of the 
spectrum. It is a consequence of the Mie scattering phase function that becomes less anisotropic at longer wavelengths, allowing a larger fraction of photons 
to escape, contributing to the increase of $P$. 

As expected from previous studies \citep{Goosmann2007,Marin2012}, the dust-induced polarization position angle is equal to 0$^\circ$ at all inclinations, 
indicating an orientation of the electric field vector perpendicular to the wind axis..

\section{Discussion}
One of the goals of \citet{Elvis2000} was to get rid of the usual, dusty torus model by introducing disk-born outflows. It is then interesting to compare
the theoretical model presented in this research note to a uniform, dusty torus scenario \citep{Kartje1995,Wolf1999,Young2000,Watanabe2003,Goosmann2007,Marin2012}. 
Spectropolarimetric modeling of circumnuclear obscuration by optically thick material showed that $P$ is expected to rise with inclination, until
the observer's line-of-sight crosses the torus horizon ($P <$ 20~\%), and then decreases down to zero at type-2 viewing angles. Moreover, the 
fraction $F$/$F_{\rm *}$ of the central flux should decrease with increasing inclination. The situation is quite different with a dusty outflow. Even if the
polarization degree behavior is approximatively reproduced with the expected perpendicular polarization angle, $P$ never exceeds 0.1~\%. 
$F$/$F_{\rm *}$ does not behave as for a dusty torus and strong fluxes are detected along the equator, which is in contradiction with observations 
\citep{Antonucci1985}.

It is also instructive to examine the difference between the structure for quasars with a flow of warm, highly ionized matter (WHIM) to its dusty 
counterpart (\citealt{Elvis2000} and Marin \& Goosmann, 2013, in press). The WHIM spectra show wavelength-independent behaviors, either in $F$/$F_{\rm *}$, 
$P$ or $\psi$. In this case, the equatorial viewing angle is found to produce a maximal polarization degree ($P \sim$ 0.3~\%) associated with a parallel photon 
position angle (i.e equals to 90$^\circ$). The intermediate WHIM inclination has a similar $P$ degree to our modeling, but both the wavelength dependence 
of $P$ and the $\psi$ angle can distinguish one model from the other.

\section{Summary and conclusions}
Using \citet{Elvis2000}'s prescriptions to characterize a disk-born outflow, we modeled the optical/UV, continuum polarization emerging from an academic 
realization of the model where the WHIM was replaced by dust. The resulting photon flux is wavelength-independent and stronger for equatorial viewing angles. 
The polarization degree spectra show wavelength-dependent signatures, increasing in the red tail, with a minimum $P$ for polar inclinations and a maximum $P$
along the outflowing direction. The overall polarization angle $\psi$ is perpendicular (i.e equal to 0$^\circ$) with respected to the projected symmetry axis
of the model. 

We thus find that a dusty, disk-born outflow shows strong differences in terms of spectropolarimetric signatures, with respect to either a regular, dusty 
torus or a WHIM scenario. Such a conclusion strengthens our previous spectropolarimetric exploration of the structure for quasars (Marin \& Goosmann, 2013, in press). 
A possible way to by-pass this problem is to consider another set of morphological parameters, where the wind has a different bending angle and/or a different 
collimation angle.

A second research note, focusing on the imaging capabilities of {\sc stokes} applied to this model will be presented. In the future, we also aim 
to push further our investigations by looking at different wavebands and computing the emission and absorption lines in different velocity fields.

\bibliographystyle{aa} 
\bibliography{marin} 

\end{document}